\documentclass{webofc}
\usepackage[varg]{txfonts}   

\usepackage{graphicx}
\usepackage{subfig}
\usepackage{color}

\begin{document}
\title{Equation of state from complex Langevin simulations}

\author{\firstname{Felipe} \lastname{Attanasio}\inst{1}\fnsep\thanks{\email{pyfelipe@thphys.uni-heidelberg.de}} 
\and
\firstname{Benjamin } \lastname{Jäger}\inst{2}\fnsep\thanks{\email{jaeger@imada.sdu.dk}} 
\and
\firstname{Felix P.G.} \lastname{Ziegler}\inst{3}\fnsep\thanks{\email{felix.ziegler@ed.ac.uk}}
}

\institute{Institute for Theoretical Physics, Universit\"{a}t Heidelberg, Philosophenweg 16, D-69120, Germany 
\and
CP3-Origins \& Danish IAS, Department of Mathematics and
Computer Science, University of Southern Denmark, Campusvej 55, 5230 Odense M,
Denmark
\and
School of Physics and Astronomy, The University of Edinburgh, EH9 3FD Edinburgh, United Kingdom
          }

\abstract{%
  We use complex Langevin simulations to study the QCD phase diagram with two light quark flavours. In this study, we use Wilson fermions with an intermediate pion mass of $\sim480\,$MeV. By studying thermodynamic quantities, in particular at lower temperatures, we are able to describe the equation of state.
}
\maketitle
\section{Introduction}\label{intro}

Determining the QCD phase diagram from first principles is a long-standing problem and has implications for heavy-ion collision experiments and cosmology. Standard lattice methods provide a non-perturbative approach for QCD, which has reached an impressive sub-percent precision. The main obstacle to direct lattice determination of the QCD phase diagram is the numerical \emph{sign} problem. For a non-vanishing chemical potential, the fermion determinant is complex, which prohibits standard importance sampling based Monte Carlo methods to be used. In recent years, a wide range of methods has been proposed to overcome or circumvent the sign problem in QCD. However, the best approach has not yet been identified. Here we present our results using the complex Langevin method~\cite{Parisi:1980ys,Parisi:1984cs}, which is based on stochastical quantisation. Over recent years, a large amount of improvements have been proposed and developed~\cite{Aarts:2008rr,Seiler:2012wz,Nagata:2015uga,Nishimura:2015pba,Nagata:2016vkn,Aarts:2017vrv,Attanasio:2018rtq,Nagata:2018net,Scherzer:2019lrh,Seiler:2020mkh}, which have made the complex Langevin method a potential contender for determining the QCD phase diagram. We report on our findings attempting an early determination of the equation of state at an intermediate pion mass. 

\section{Setup}

We use a lattice setup with two mass-degenerate dynamical quarks. The lattice parameters are: the hopping parameter is $\kappa = 0.1544$ and the gauge coupling is $\beta = 5.8$. This implies a fixed-scale approach. These parameters are based on the simulations of~\cite{DelDebbio:2005qa}. Our determination of the lattice spacing, performed via the Wilson flow, has lead to $a\sim0.06\,\mathrm{fm}$, consisten with the~\cite{DelDebbio:2005qa}. We use a spatial volume of $24^3$ in lattice units. To map the phase diagram, we vary the temporal extent from $4$ up to $128$ sites, but most of our simulations are in the region of $N_t \leq 32$. The temperature range covers $822$ to $25\,$ MeV. For the quark chemical potential, we use values $a \mu = 0 - 2$, in physical units up to $\sim6000\,$MeV. To overcome runaway solutions and instabilities, we use adaptive step size scaling. In addition, we used one step of gauge cooling after every Langevin step. For this setup, we combined it with dynamic stabilisation (DS) with $\alpha_{\mathrm{DS}} = 1$. Here we employ a modified version of DS for which we do not use the adaptive step size scaling in the additional DS drift; instead, we keep the step size fixed in order to ensure that the additional force term only depends on the distance to SU($3$), i.e.,
\begin{equation}
    \overline{\epsilon} \, \alpha_{DS} \, K_{DS}.
\end{equation}
The step size of the dynamic stabilization force is fixed to $\overline{\epsilon} = 10^{-3}$. For our simulations, we use a modified version of the openQCD v1.6 \cite{openqcd}
and openQCD-fastsum software packages \cite{openqcd-fastsum}.   
\begin{figure}
\centering
\subfloat{
  \includegraphics[width=65mm]{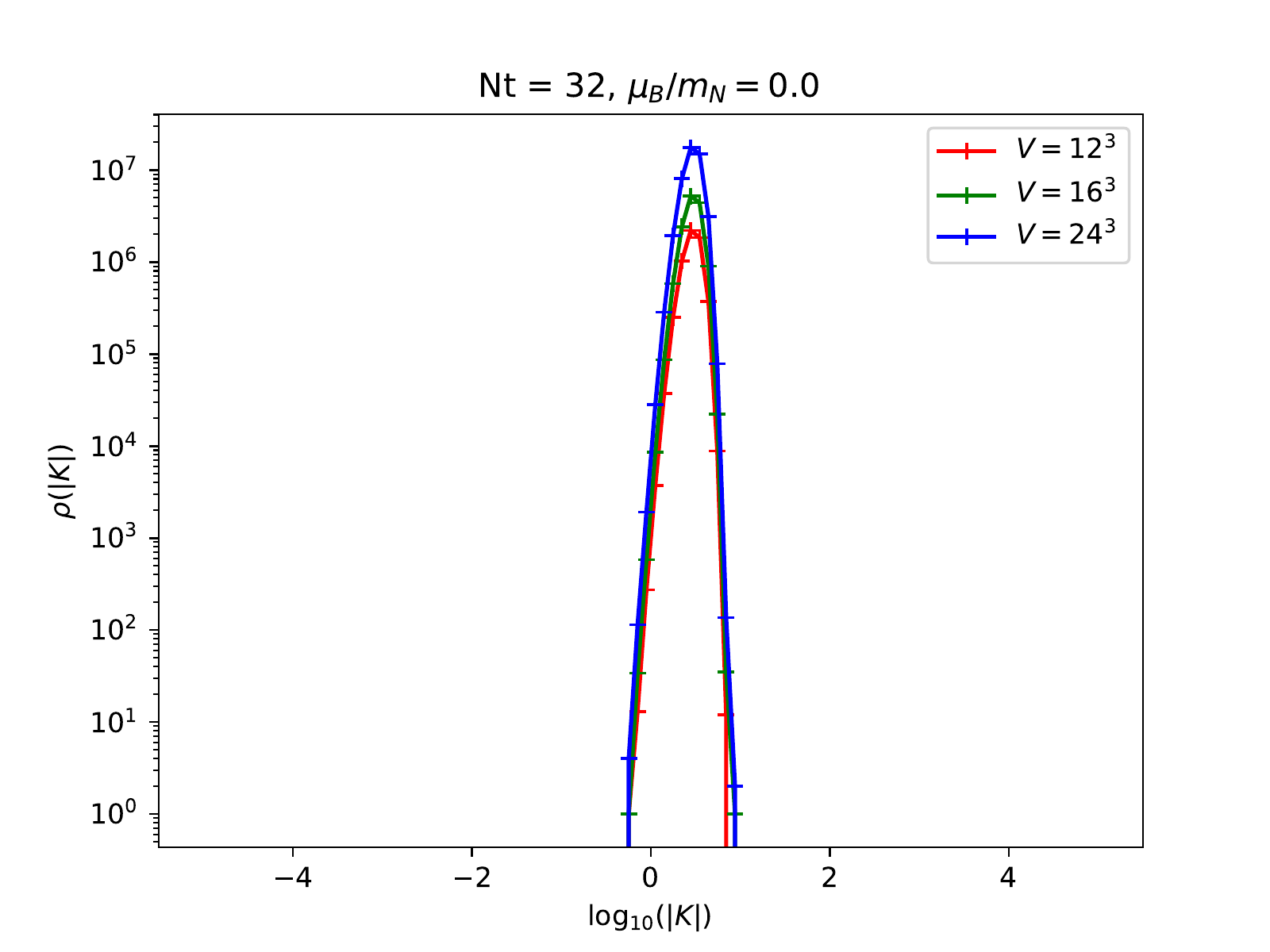}
}
\subfloat{
  \includegraphics[width=65mm]{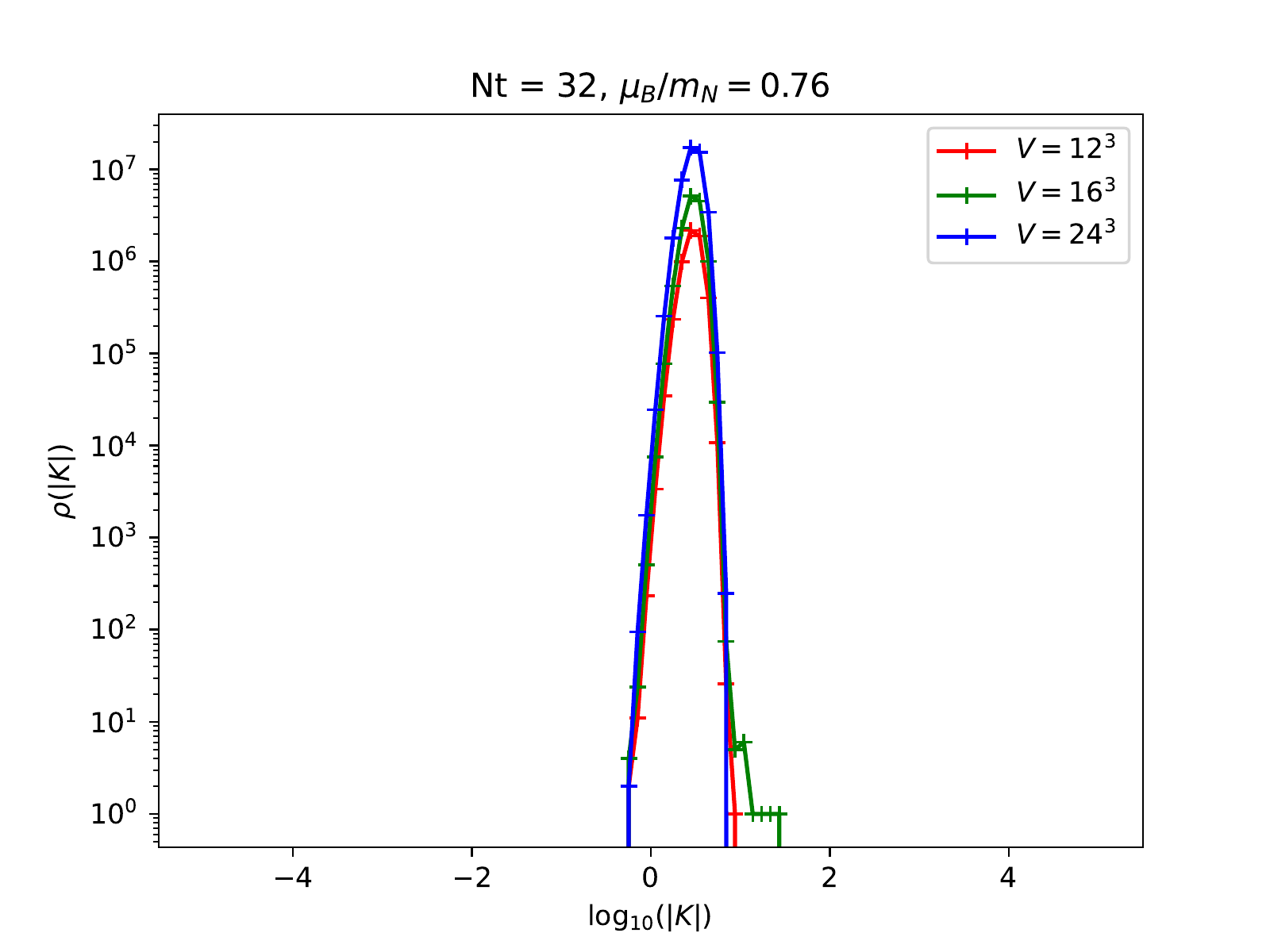}
}
\hspace{0mm}
\subfloat{
  \includegraphics[width=65mm]{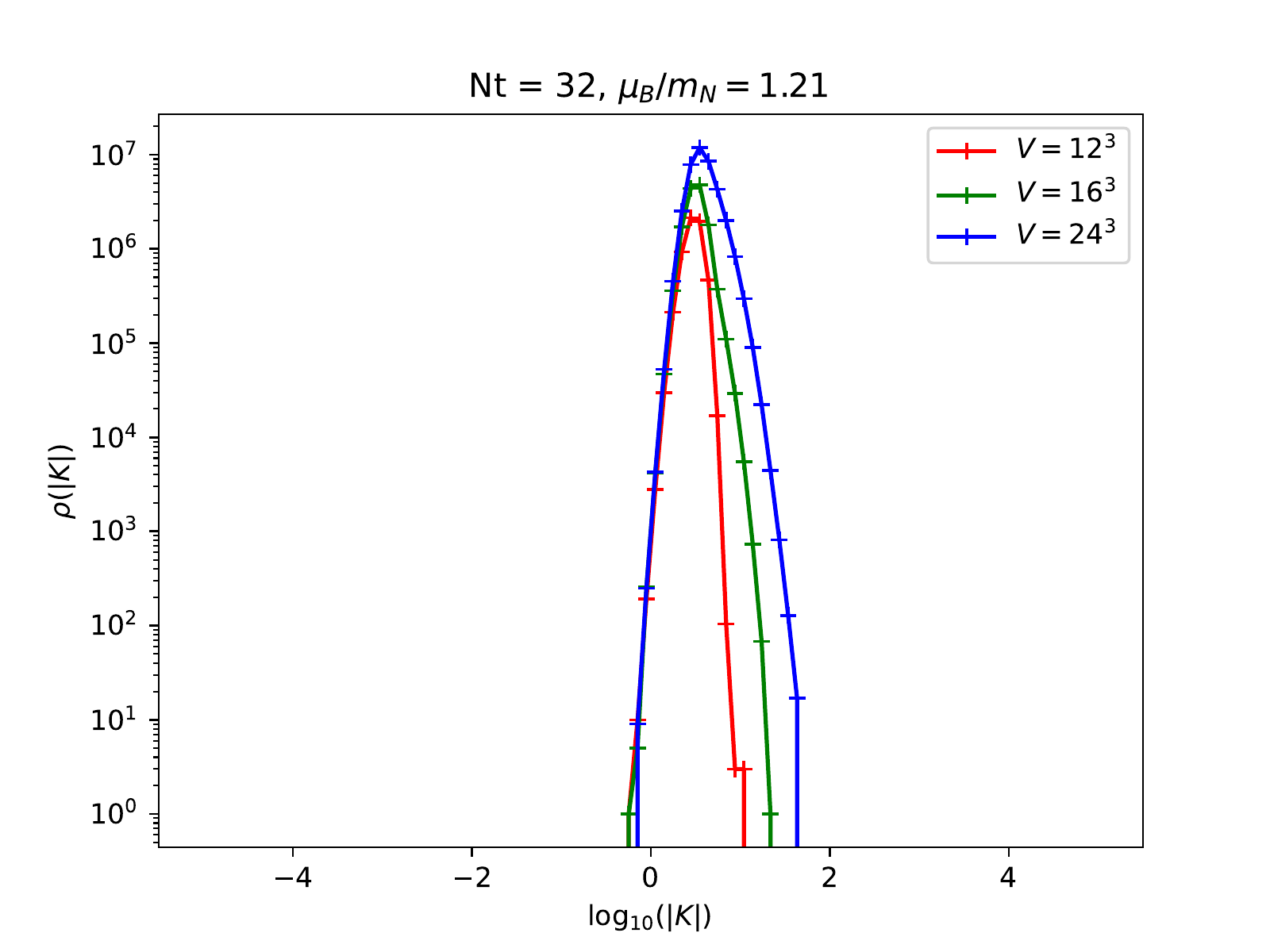}
}
\subfloat{
  \includegraphics[width=65mm]{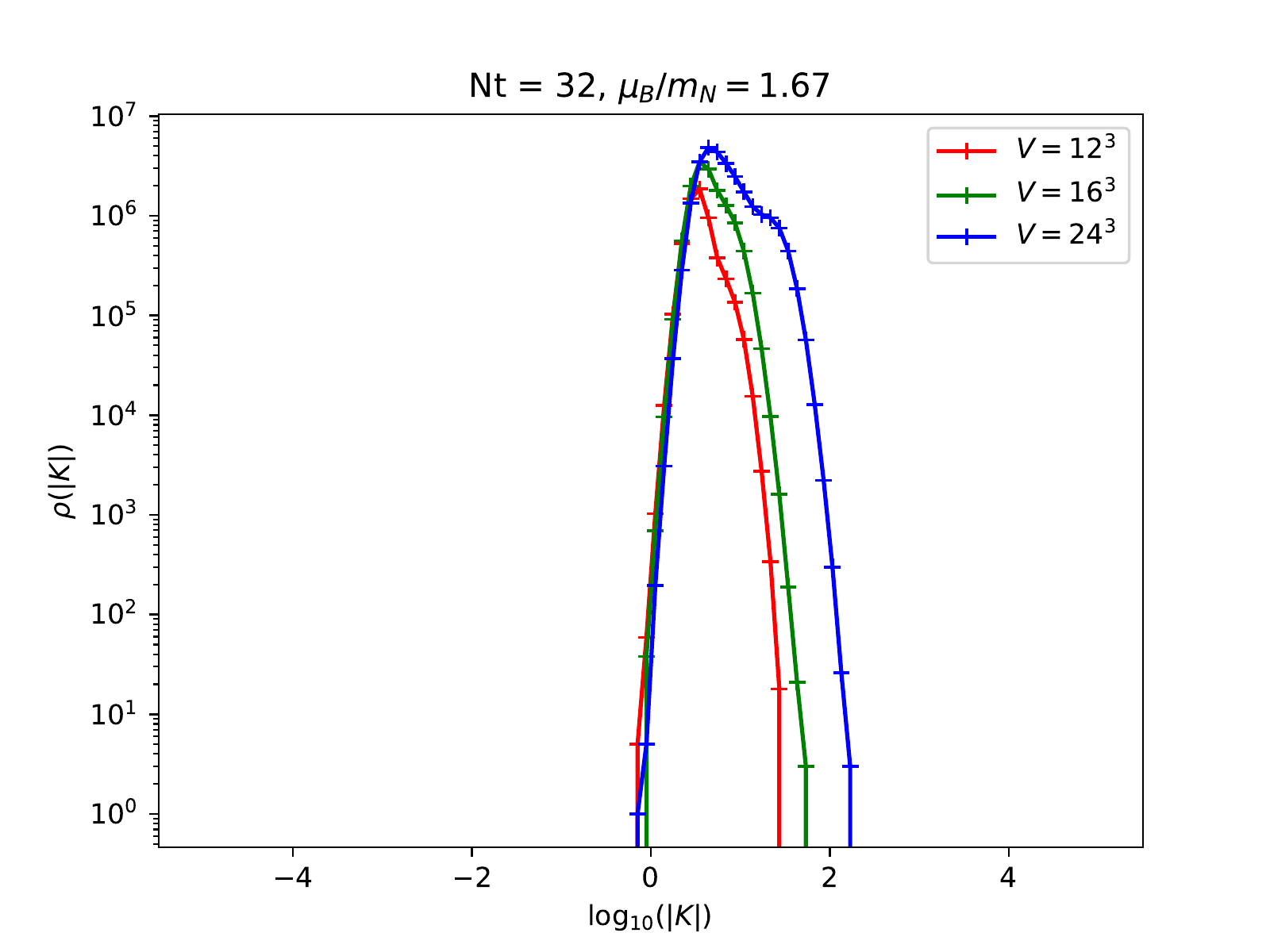}
}
\caption{Histograms of the Langevin force $K$ for $N_t=32$ ($T = 103\,$MeV) for three volumes and different values of the baryon chemical potential $\mu_B$ in units of the nucleon mass.}
\label{fig.hist}
\end{figure}
As an indication of the stability of the simulation, we show histograms of the Langevin 
force for various baryon chemical potentials\footnote{The baryon chemical potential $\mu_B$ is 
related to the quark chemical potential $\mu$ by: $\mu = \mu_B / 3$.}, which can be seen in Figure~\ref{fig.hist}. These results were also obtained at 
smaller volumes, i.e. $16^3$ and $12^3$. We find that all histograms show a localised 
distribution, which typically is taken as an indication for correct convergence of the CL
method
\cite{Nagata:2016vkn}. With increasing $\mu$ the histograms show some structure emerging, 
which indicates that the chemical potential is changing the Langevin forces. However, the 
histograms or skirts remain sufficiently compact over the entire range.

\section{Results}

To identify the phase boundaries, we study the Polyakov loop and the fermion density. 
\begin{figure}[h!]
\centering
\subfloat{
  \includegraphics[width=65mm]{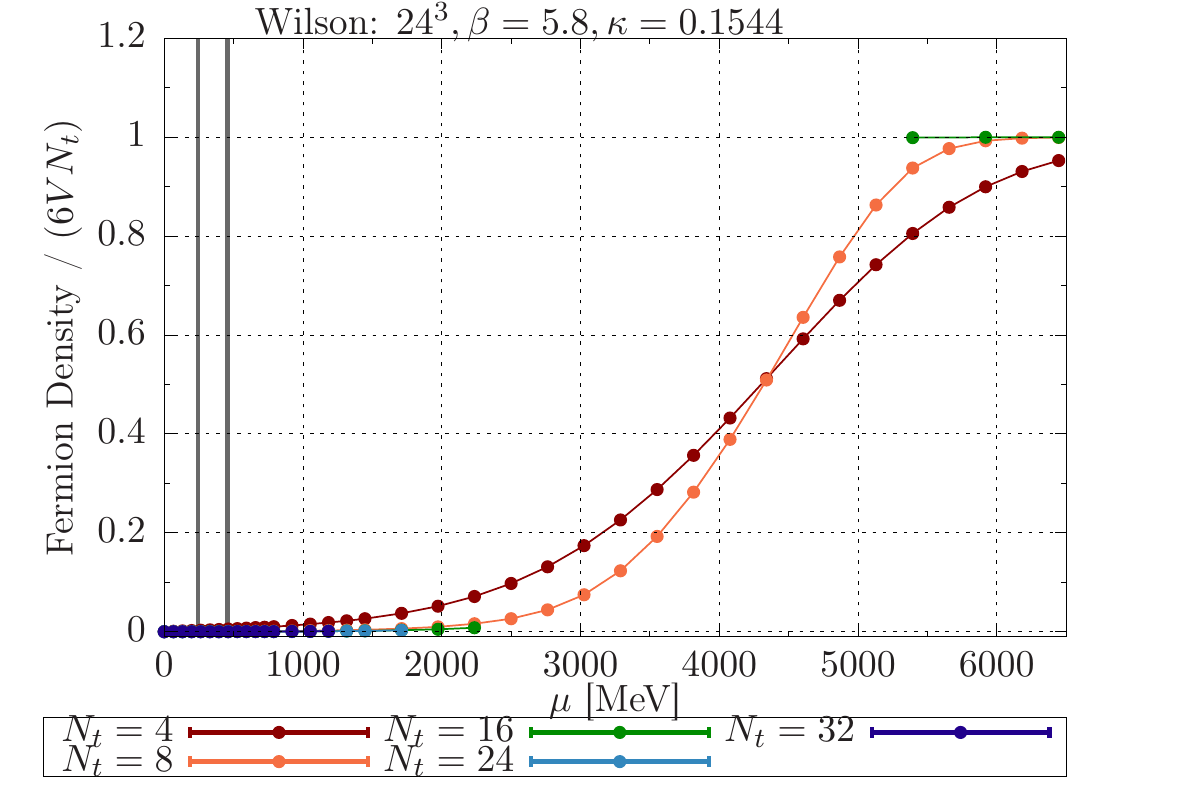}
}
\subfloat{
  \includegraphics[width=65mm]{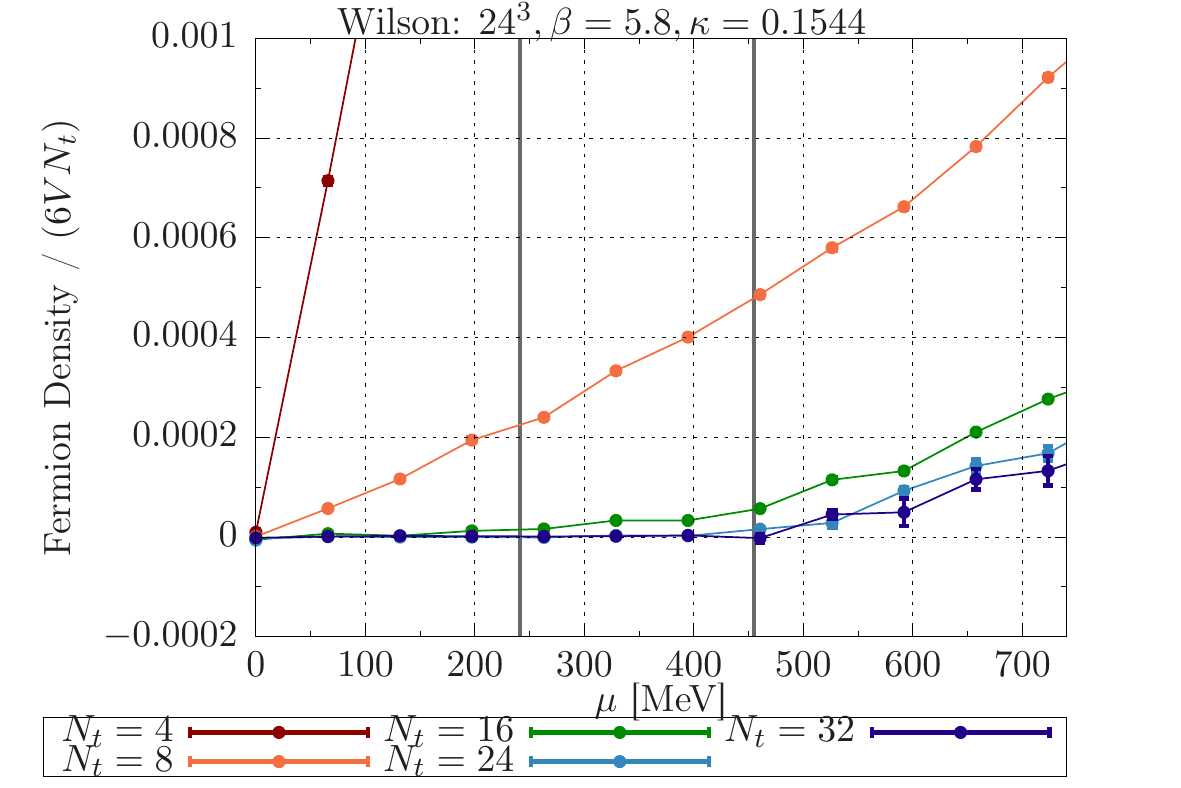}
}
\caption{Fermion density as a function of the quark chemical potential. The vertical grey lines correspond to $m_\pi/2$ and  $m_N/3$ respectively.}
\label{fig.dens}
\end{figure}
Figure~\ref{fig.dens} shows the fermion density as a function of the quark chemical potential. The plot on the left shows the entire simulation range. We found that the density saturates as expected for a large enough density. This is a lattice artifact that appears if all lattice sites are filled with $N_{\mathrm{spin}} \times N_c = 6$ fermionic degrees of freedom per quark flavour. Adding more is not possible because of Pauli blocking. On the right hand side of Figure~\ref{fig.dens} we show an enlarged view, which highlights the region up to $700\,$MeV. As expected, colder simulations show a later rise of the density. For the lowest temperature shown here, we see that density remains approximately zero until $m_N/3$, where we expect that quarks can be created. This phenomenon is also known as the Silver Blaze and emerges dynamically.  
\begin{figure}
\centering
\subfloat{
  \includegraphics[width=65mm]{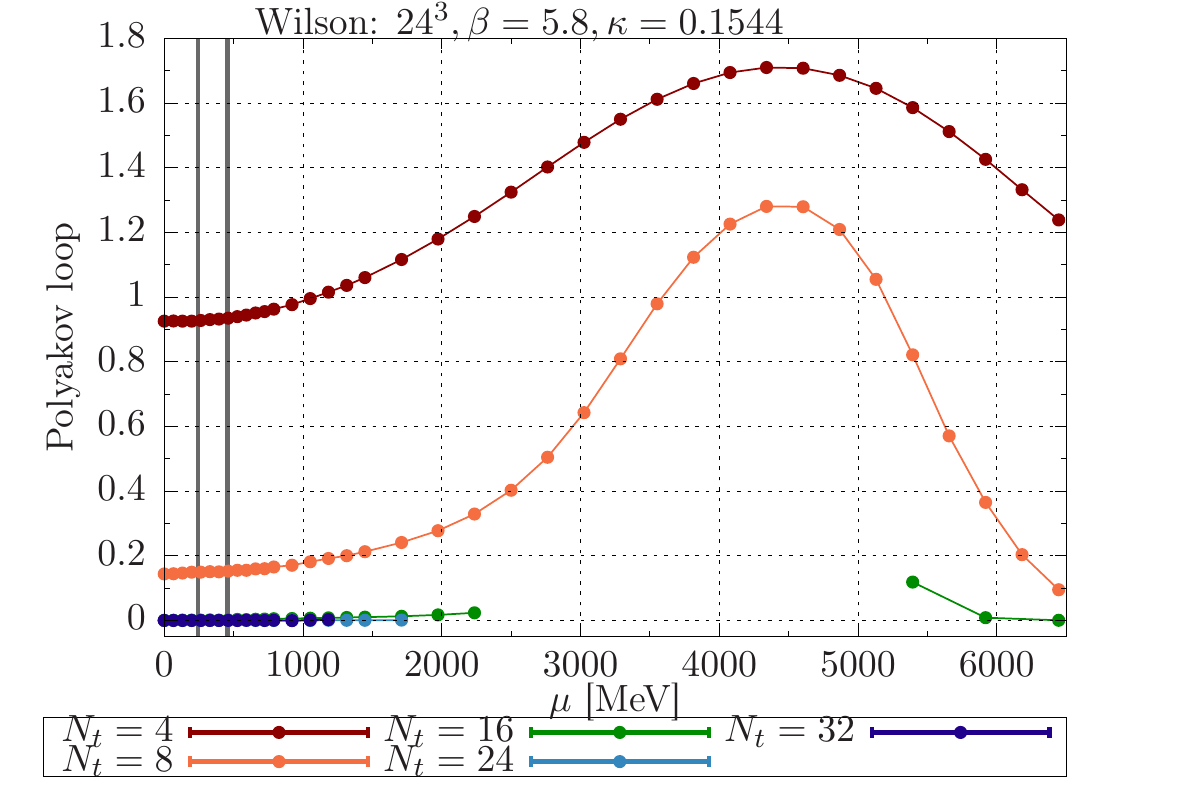}
}
\subfloat{
  \includegraphics[width=65mm]{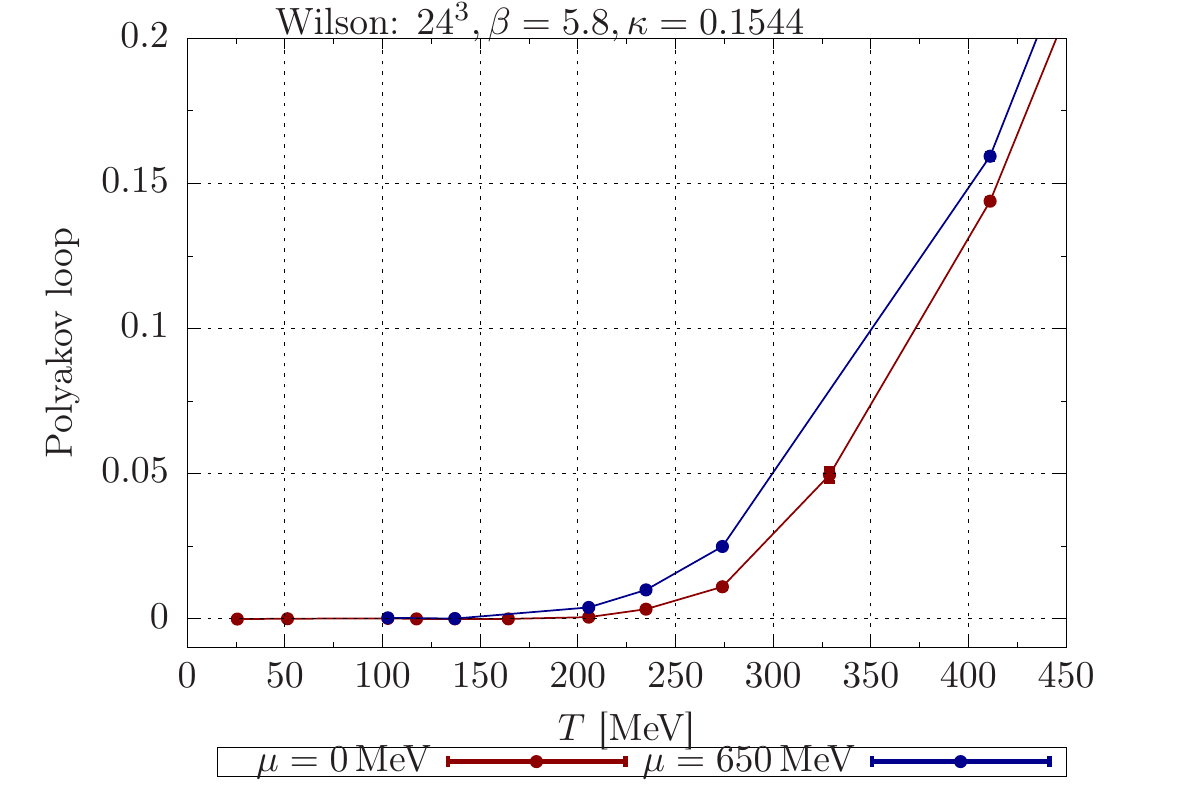}
}
\caption{Polyakov loop as a function of the quark chemical potential (left) and temperature (right). The saturation density is reached at $\mu \sim 4000\,$MeV.}
\label{fig.poly}
\end{figure}
Figure~\ref{fig.poly} shows the Polyakov loop. On the left-hand side as a function of the chemical potential. The same lattice artefact as seen in the density is visible. However, here it causes the Polyakov loop to shrink again. The right panel of Figure~\ref{fig.poly} provides a different perspective on the transition. For larger chemical potentials, the transition appears to occur at lower temperatures. This also is in line with our expectations that the transition temperature reduces as the chemical potential increases. We fit the fermion density to a cubic polynomial and use the result to compute the pressure and from it the equation of state.
\begin{figure}
\centering
\subfloat{
  \includegraphics[width=65mm]{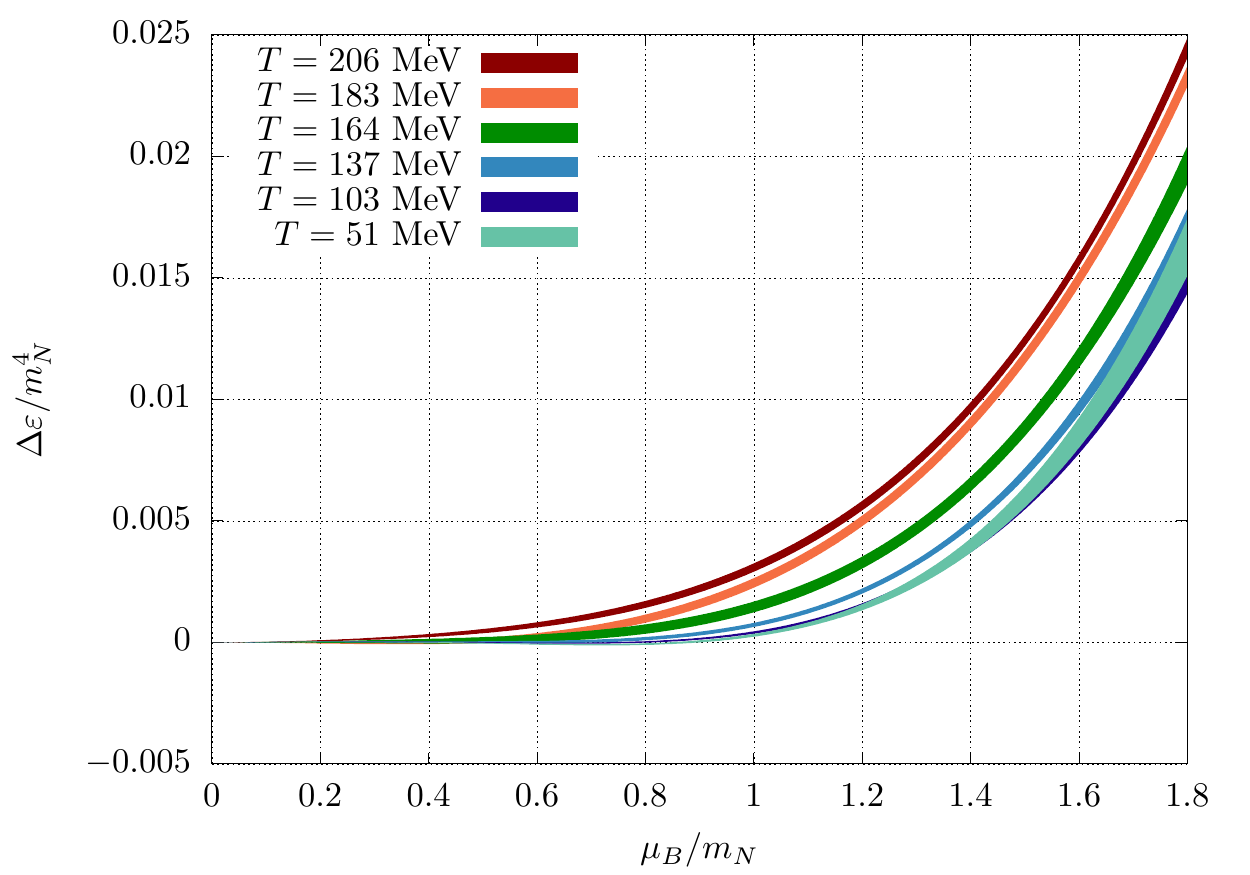}
}
\subfloat{
  \includegraphics[width=65mm]{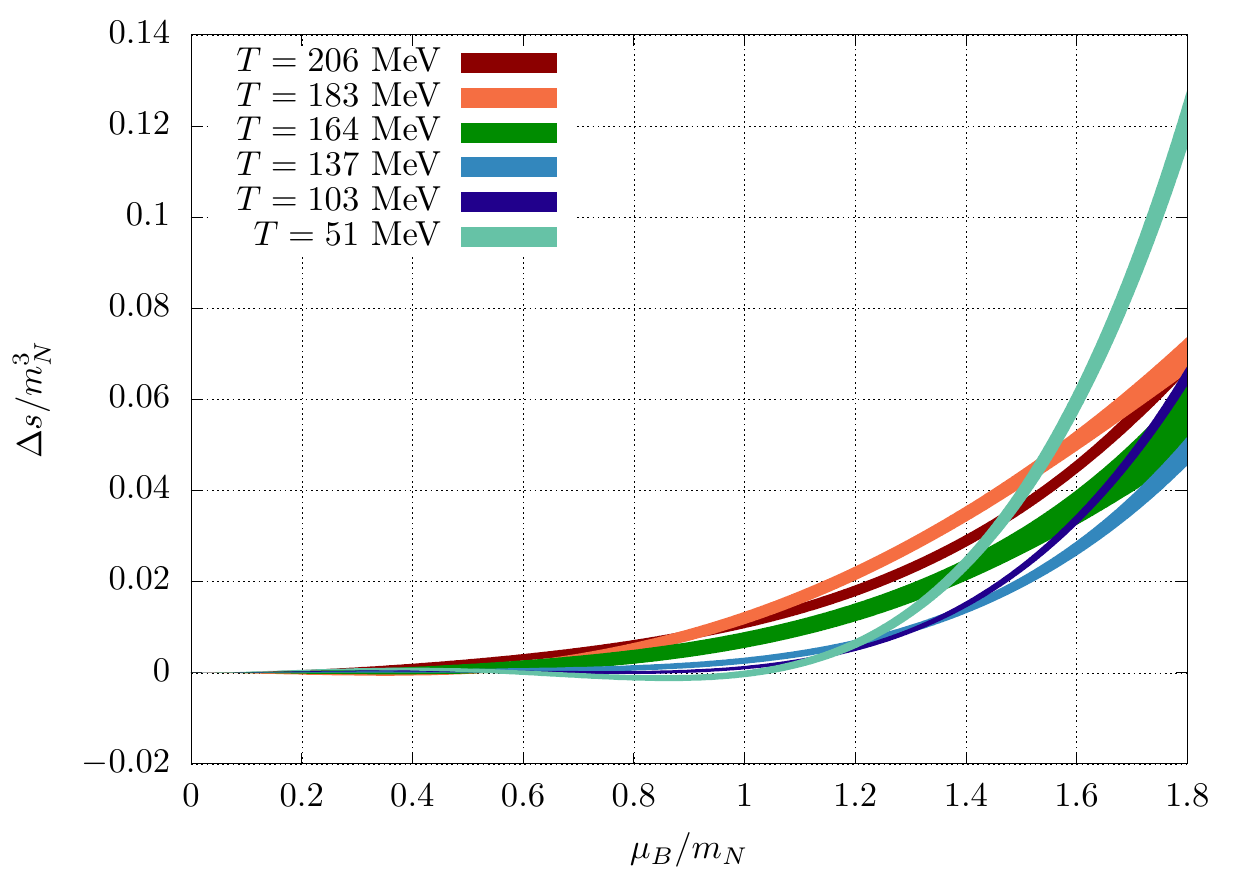}
}
\caption{The equation of state extracted from the fermion density.}
\label{fig.eos}
\end{figure} 
Figure~\ref{fig.eos} shows the energy and entropy density, respectively. We find that the equation of state becomes stiffer with decreasing temperatures.

\section{Conclusion and Outlook}

Here we presented an update on our efforts to determine the equation of state from complex Langevin simulations. Although our simulations are still at an unphysical quark mass, which corresponds to a pion mass of roughly $480\,$MeV, our simulations confirm the expectations on the QCD phase diagram. In the future, we will improve our simulation by using $\mathcal{O}(a)$ improved actions. We hope that this allows us also to reduce the pion mass closer to the physical world. In addition, we plan to study multiple volumes to get a handle on the order of the transition(s). 

\section{Acknowledgement}

Computing resources were provided by the University of Southern Denmark, Deic initiative (UCloud, GenomeDK) and PRACE on Hawk (Stuttgart) with project ID 2018194714. This work used the DiRAC Extreme Scaling service at the University of Edinburgh, operated by the Edinburgh Parallel Computing Centre on behalf of the STFC DiRAC HPC Facility (www.dirac.ac.uk). We are thankful for the support of the  Deutsche Forschungsgemeinschaft (DFG) under EXC2181/1 - 390900948 (STRUCTURES) and the SFB 1225 (ISOQUANT).

\bibliography{references}

\end{document}